\documentclass[12pt]{iopart}

\usepackage{graphicx}
\usepackage{iopams}
\usepackage{setstack}
\usepackage{subfigure}

\newcommand{\pd}[2]{\frac{\partial #1}{\partial #2}}

\newcommand{\link}{\prec\!\!*\,}

\begin{document}

\title{Particle simulations in causal set theory}
\author{L Philpott}
\address{Blackett Laboratory, Imperial College, London SW7 2AZ, UK}
\ead{l.philpott06@imperial.ac.uk}

\begin{abstract}
Models of particle propagation in causal set theory are investigated through simulations. For the swerves model the simulations are shown to agree with the expected continuum diffusion behaviour. Given the limitations on the simulated causal set size, the agreement is far better than anticipated.
\end{abstract}

\pacs{04.60.Bc}

\section{Introduction}

Causal set theory is a discrete, Lorentz invariant approach to quantum gravity. 
For comprehensive reviews of the field see, for example~\cite{Henson:2006kf, Dowker:2006sb, Sorkin:2003bx}. The phenomenology of particles in causal set theory was first investigated by Dowker et al.~\cite{Dowker:2003hb}, and later further developed in~\cite{Philpott:2008vd}. Massive particles propagating in a discrete spacetime are expected to experience small fluctuations in momentum, an effect termed `swerves'. Several simple classical microscopic models for particle propagation on a causal set were proposed in~\cite{Dowker:2003hb} and~\cite{Philpott:2008vd}, but it was shown that it was not necessary to choose a specific model to understand the phenomenology in the continuum limit. Dowker et al.~demonstrated that any Lorentz invariant Markovian stochastic process on the massive particle state space gives rise to a continuum diffusion equation. In terms of the observers time parameter, `cosmic time', the swerves diffusion equation is~\cite{Philpott:2008vd}:
\begin{equation}
\pd{\rho}{t} = \frac{-p^i}{m\gamma}\partial_i\rho + k\partial_a\left(g^{ab}\sqrt{g}\partial_b\left(\frac{\rho}{\gamma\sqrt{g}}\right)\right)\,,
\end{equation}
where $\rho$ is the probability density for the system to have a certain momentum and position at a given time, $p$ is the particle momentum, $m$ is the particle mass, $\gamma = \sqrt{m^2+p^2}/m$ is the standard relativistic factor, and $g$ is the metric on the mass shell, i.e.~the hyperboloid in momentum space defined by $p_{\mu}p^{\mu}=-m^2$. $k$ is the diffusion parameter, the one free parameter of the model. 

Although no one microscopic model was chosen to derive the diffusion equation, the phenomenological parameter, $k$, will in reality depend on the properties of the underlying model for particles on a causal set. This paper investigates whether the behaviour of underlying particle models is well described by the continuum limit swerves diffusion equation.
As will be shown below, the results are in fact better than expected, the model proposed in~\cite{Dowker:2003hb} is well approximated by the diffusion equation even far from the continuum limit. This allows the relationship between the microscopic model parameters and the diffusion parameter to be determined.

First it is necessary to review the basic concepts of causal set theory. A causal set is a set $C$ endowed with a binary relation `precedes', $\prec$, that satisfies:
\begin{enumerate}
        \item transitivity: if $x\prec y$ and $y\prec z$ then $x\prec z$, $\forall x,y,z\in C$;
        \item reflexivity: $x\prec x$, $\forall x \in C$;
        \item acyclicity: if $x\prec y$ and $y \prec x$ then $x=y$, $\forall x, y \in C$;
        \item local finiteness: $\forall x, z\in C$ the set $\left\{y\mid x\prec y\prec z\right\}$ of elements is finite.
\end{enumerate}

Some definitions will be useful for the following work.
Let $C$ be a causal set.
\begin{enumerate}
        \item A \textit{chain} is a totally ordered subset of $C$.
        \item A \textit{longest} chain between two elements $x,\;y\in C$ is a chain whose length is longest amongst chains between those endpoints. There may be more than one longest chain between two elements. The length of the longest chain between elements $x,\;y\in C$ will be denoted $d(x,y)$. 
        \item A \textit{link} is an irreducible relation: elements $x$ and $y$ are linked if and only if $d(x,y)=1$. If two elements $x,\;y\in C$ are linked, it will be denoted $x\link y$.
        \item A \textit{path} is a chain consisting of links.  
\end{enumerate}

\section{Particle models}

For the purpose of simulations, a causal set can be constructed from a continuum manifold by a process called sprinkling. Points are selected at random from a manifold via a Poisson process in which the probability measure is equal to the spacetime volume measure in fundamental units. The causal order on the points induces the partial order on the elements in the causal set.

Construct a causal set by sprinkling into Minkowski spacetime. A massive particle trajectory is taken to be a chain of elements $e_n$ in the causal set, i.e.~a linearly ordered subset of $C$. The particle trajectory is constructed iteratively. It is assumed that the trajectory's past determines its future, but that only a certain amount of the past is relevant.

\subsection{Swerves}
First proposed in~\cite{Dowker:2003hb}, this model relies on information about the approximating spacetime. Suppose the particle is currently located `on' an element $e_n$, with a four-momentum $p_n$. The next element, $e_{n+1}$ is chosen such that
\begin{itemize}
\item $e_{n+1}$ is in the causal future of $e_n$ and within a propertime $\tau_f$ of $e_n$,
\item the momentum change $|p_{n+1}-p_n|$ is minimized.
\end{itemize}
$\tau_f$ is the forgetting time of the process. The momentum $p_{n+1}$ is defined to be proportional to the vector between $e_n$ and $e_{n+1}$, normalized by the particle mass $m$.

\subsection{Intrinsic models}
Two models intrinsic to the causal set, i.e.~not relying on continuum information, were proposed in~\cite{Philpott:2008vd} to illustrate the range of possibilities available. Rather than a forgetting time, these models depend on a forgetting number $n_f$.

\subsubsection{Model 1}
 Given a partial particle trajectory $\ldots e_{n-1},e_n$ the next element $e_{n+1}$ is chosen such that
\begin{itemize}
\item $d(e_{n-1},e_{n+1})\leq 2n_f$,
\item $d(e_n,e_{n+1})$ is maximized subject to $d(e_n,e_{n+1})\leq n_f$.
\end{itemize}
These requirements do not guarantee the existence of a unique $e_{n+1}$. There will, however, almost surely be finitely many eligible elements and the trajectory can be constructed by choosing an element uniformly at random from these. Note that this model is slightly different from the first intrinsic model given in~\cite{Philpott:2008vd}, where equalities in the above conditions were given. Model 1 of~\cite{Philpott:2008vd} does not guarantee the existence of an $e_{n+1}$ under reasonable conditions. 

\subsubsection{Model 2}
The trajectory is constructed as a path in this model, i.e. $d(e_n,e_{n+1})=1$ for any $e_n$, $e_{n+1}$. 
Given a partial particle trajectory $\ldots e_{n-n_f}, \ldots, e_{n-1},e_n$ the next element $e_{n+1}$ is chosen such that
\begin{itemize}
\item $d(e_n,e_{n+1})=1$,
\item $d(e_{n-n_f},e_{n+1}) + \ldots + d(e_{n-1},e_{n+1}) + d(e_n, e_{n+1})$ is minimized
\end{itemize}
Again this minimization does not necessarily yield a unique $e_{n+1}$, in which case the trajectory is constructed by choosing an element uniformly at random from those eligible. Also, if the trajectory has length less than $n_f$ the minimization is done over all elements available. 

In these models it is assumed that the forgetting parameter $\tau_f$ or $n_f$ is many orders of magnitude greater than the discreteness scale.

\section{Numerical results}

\begin{figure}[t]
\begin{center}
\subfigure[Model 1, $N=32768$, $\tau_f = 0.1105$, $l=16$.]{
\includegraphics[width= 0.45\textwidth]{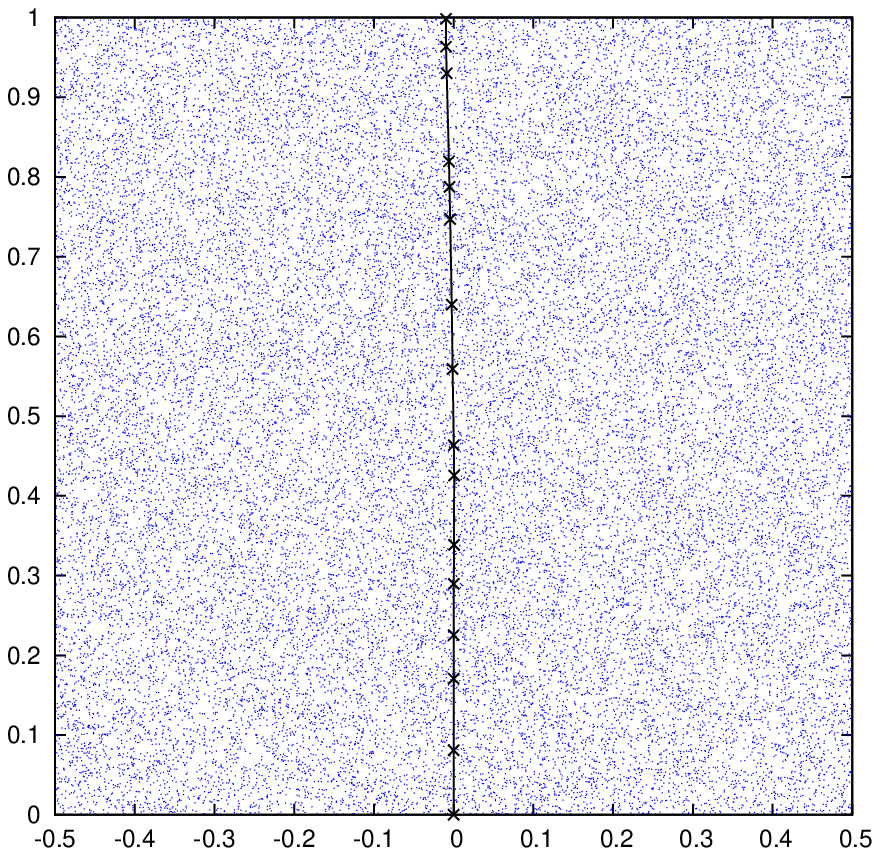}
\label{f:2Dexamplemom}}
\subfigure[Model 2, $N=32768$, $n_f = 20$, $l=13$.]{
\includegraphics[width= 0.45\textwidth]{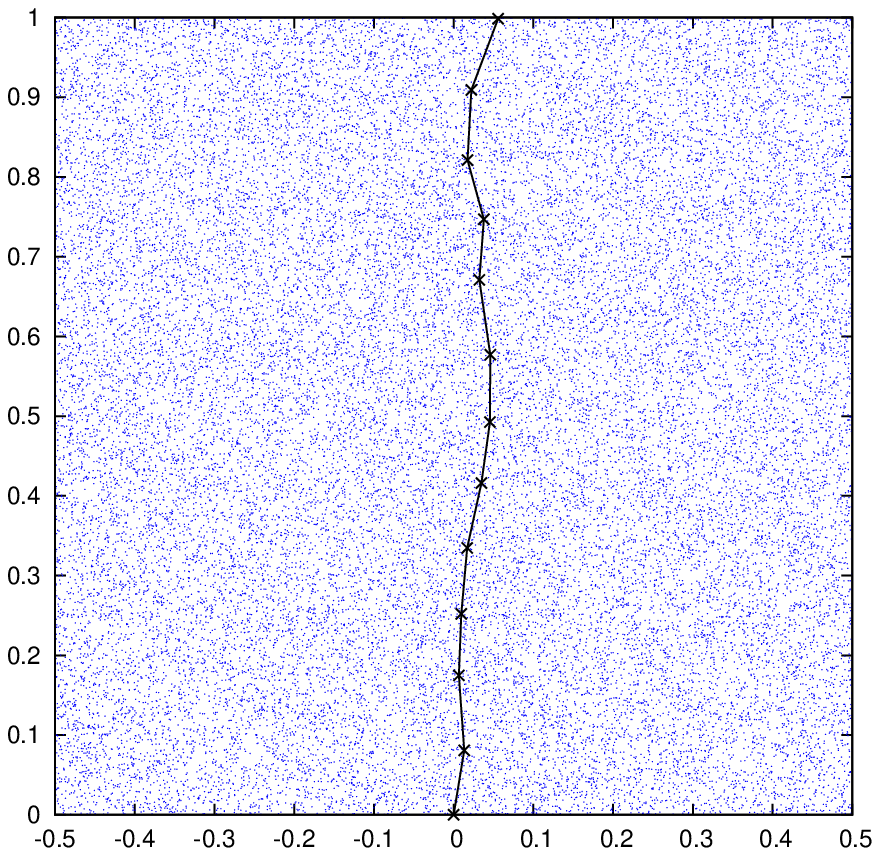}
\label{f:2Dexampleintrinsicnf}}
\subfigure[Model 3, $N=32768$, $n_f = 20$, $l=163$.]{
\includegraphics[width = 0.45\textwidth]{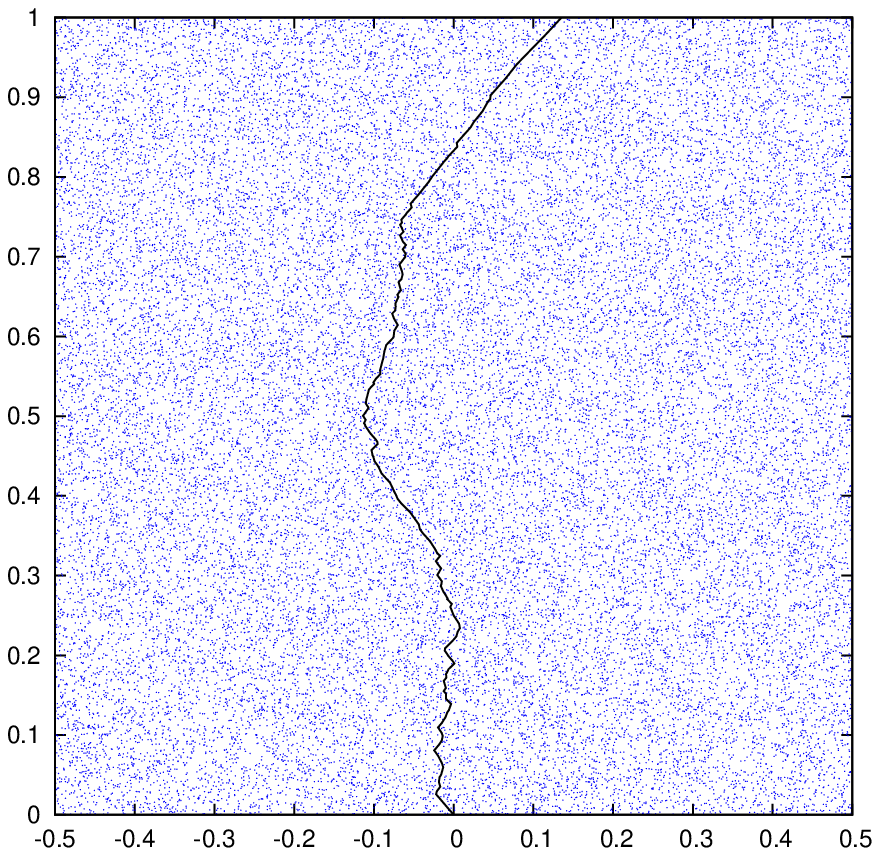}\label{f:2Dexampleintrinsic}}
\caption{Example trajectories with causal set size, $N=32768$, forgetting parameter $\tau_f/n_f$ and trajectory length $l$.}\label{f:exampletraj}
\end{center}
\end{figure}

Simulations of the particle models given above were developed within the Cactus numerical relativity framework~\cite{Cactus}, making use of the CausalSets arrangement written by David Rideout. Although the models are entirely general, due to computational limitations simulations were carried out in 1+1 dimensions. Points were sprinkled into a region of Minkowski spacetime with a Poisson distribution with a mean number of elements $N$. One additional point was added to each causal set at the origin to give a fixed beginning point, $e_1$, for the trajectories. For convenience, the particular frame to which the coordinates of the sprinkled points in Minkowski spacetime refer will be called the embedding frame.  For the swerves model the particle was assumed to be initially at rest in the embedding frame, i.e.~$p_1=(m,0)$. For the intrinsic models the particles were also assumed to be initially close to rest. To construct the first step for the intrinsic model 1, the condition $d(e_{0},e_{2})\leq 2n_f$ was neglected and $e_2$ was taken to be the first (in time) element that maximized $d(e_1,e_2)\leq n_f$. In the case of intrinsic model 2 an additional point was added to the causal set at $x=0,\, t=0.25$ and the beginning of the trajectory was taken to be a longest chain between the points $x=0,\, t=0$ and $x=0,\, t=0.25$, to ensure the particle was initially close to rest. An example trajectory for each of the models is shown in Figure~\ref{f:exampletraj}. Each model gives fluctuations in momentum, but the amount of fluctuation is clearly model dependent. Note that for a 1+1-dimensional causal set with $N=32768$, $n_f=20$ is roughly equivalent to $\tau_f=0.1105$ if we take $d_{pl}=\sqrt{V/N}$ and $n_f=\tau_f/d_{pl}$.

To investigate how well the diffusion equation approximates the microscopic model, and the relationship between the phenomenological parameter and the underlying model parameters, we focused on the swerves model. For a given causal set and initial position and momentum, the swerves model defines a unique trajectory. To say that the swerves model results in diffusion is, in a sense, saying that the exact underlying causal set is unknown. To simulate this, many different sprinklings into the same region of Minkowski spacetime were generated, and the unique trajectory in each was calculated. For each trajectory, the `final' position and momentum were determined as the trajectory crossed the $t_f=0.95$ hypersurface. This gave a distribution in position and momentum that could be compared to that expected from the 1+1 dimensional swerves diffusion equation:
\begin{equation}
\pd{\rho}{t} = \frac{-p}{\sqrt{m^2+p^2}}\pd{\rho}{x} + k\pd{}{p}\left(\frac{\sqrt{m^2+p^2}}{m}\pd{\rho}{p}\right)\,.
\label{e:diffeqn}
\end{equation}

Before comparing the simulation results and the diffusion equation we must note that some trajectories may need to be rejected. If a trajectory is close to the boundary of the region of Minkowski spacetime at any point, it will `bounce' back and distort the results. For the results shown here, the models parameters were chosen such that there were few such `invalid' trajectories -- any that did occur were removed from the results.
 
The microscopic model contains three parameters: the forgetting parameter, $\tau_f$, a discreteness scale, $d_{pl}$, and the particle mass, $m$. Although the discreteness scale does not appear in the trajectory algorithm, the trajectory clearly depends on how many points have been sprinkled into a given volume. For the purposes of this investigation the discreteness length can be defined as $d_{pl} = \sqrt{V/N}$, where $V$ is the volume of the region of 1+1 dimensional Minkowski spacetime and $N$ is the mean number of causal set elements sprinkled. 

The particle mass appears in the swerves model algorithm only to normalize the momentum at each step. The trajectory constructed is, in fact, independent of the mass. The final momentum distribution from a collection of trajectories does depend on the mass, but changing the mass only rescales the momentum. Thus, without running any simulations it is possible to determine the dependence of the diffusion parameter, $k$, on $m$: examining (\ref{e:diffeqn}), if $m$ and $p$ are rescaled by a factor $\alpha$ then $k$ must be rescaled by $\alpha^2$, therefore $k\sim m^2$. 

To determine the dependence of $k$ on $\tau_f$, 500 trajectories were evolved for each of 11 values of $\tau_f$ between 0.03 and 0.1, with $m = 1$ and $d_{pl}\sim 0.0055$, in embedding units. For each value of $\tau_f$ final position and momentum histograms were calculated. The swerves diffusion equation was numerically evolved for a range of values of $k$ and a best fit value of $k$ was determined for each $\tau_f$ by minimizing the reduced $\chi^2$ value
\begin{equation}
\chi^2_{red} = \frac{1}{f}\sum_i\frac{\left(O_i-E_i\right)^2}{E_i}\,,
\end{equation}
where $O_i$ is the observed frequency for a momentum bin $i$, $E_i$ is the expected frequency (i.e.~that given by the evolution of the diffusion equation), and $f$ is the number of degrees of freedom (here, $f=$~number of data points - 1). $\chi^2$ is not a good measure of fit if a significant proportion of the expected frequencies are less than five (see, e.g.~\cite{Sheskin:2004}). To avoid this problem multiple bins were combined where necessary. A reduced $\chi^2$ of order 1 is usually considered to indicate a good fit (see, e.g.~\cite{Taylor:1997}). The momentum diffusion was chosen for the comparison as the position diffusion is driven by the momentum diffusion. 

\begin{figure}[t]
\begin{center}
\subfigure[Position]{
\includegraphics[width = 0.45\textwidth]{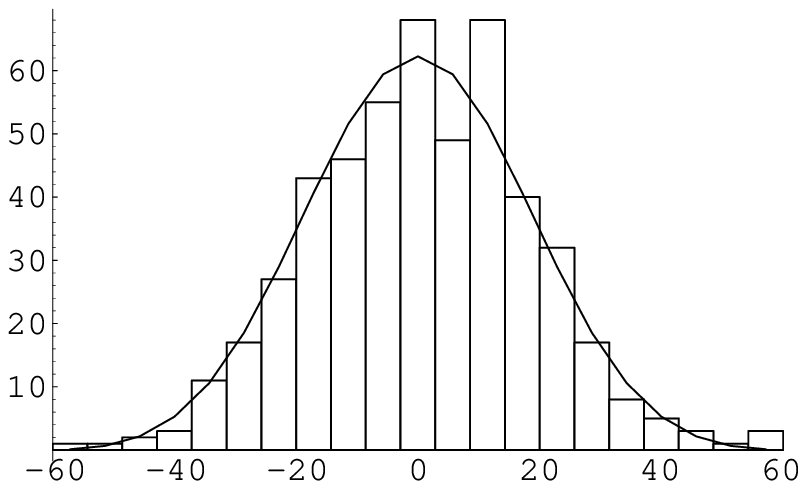}}
\subfigure[Momentum]{
\includegraphics[width = 0.45\textwidth]{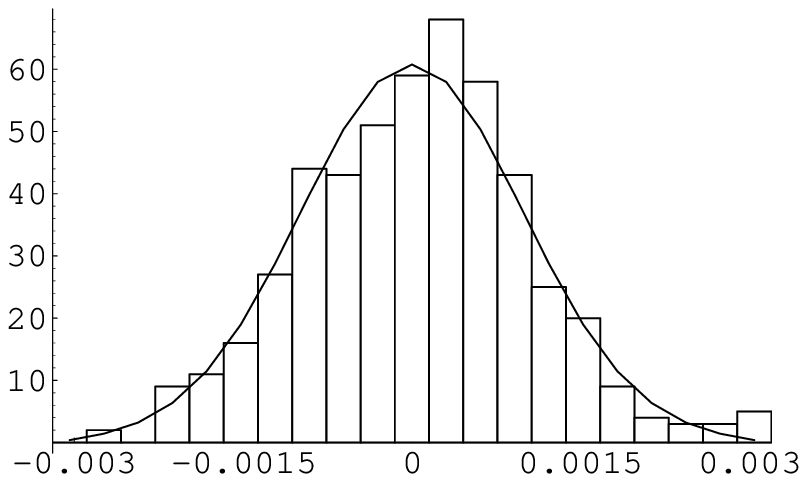}}
\caption{Histograms for 500 trajectories with $\tau_f=0.04$, $N=32768$ and best fit solutions from the swerves diffusion equation, $k=2.8\times 10^{-9}$. The reduced $\chi^2$ value for the momentum distribution is $\chi^2_{red}=0.59$.}\label{f:histograms}
\end{center}
\end{figure}

Example position and momentum histograms and the corresponding best fit solutions for $\tau_f=0.04$ are shown in Figure~\ref{f:histograms}. Note that $\tau_f=0.04$ has not been chosen for any particular reason, the other values of $\tau_f$ have equally good fits. It is clear from this figure that the swerves diffusion equation is a very good approximation to the underlying model even though $\tau_f$ is not many orders of magnitude greater than $d_{pl}$ and the discreteness length is nowhere near the $d_{pl}\rightarrow 0$ continuum limit. 

\begin{figure}[p]
\begin{center}
\subfigure[$\ln k$ vs.~$\ln\tau_f$ and best fit line $-5.02\ln\tau_f-9.8$]{
\hspace{-0.05\textwidth}
\includegraphics[width = 0.554\textwidth]{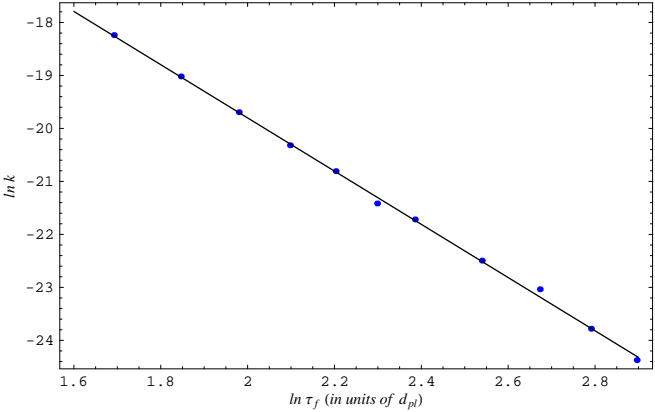}
\label{f:logkvslogtauf}}
\subfigure[$\ln k$ vs.~$\ln d_{pl}$ and best fit line $4.0\ln d_pl + 13$]{
\hspace{-0.04\textwidth}
\includegraphics[width = 0.546\textwidth]{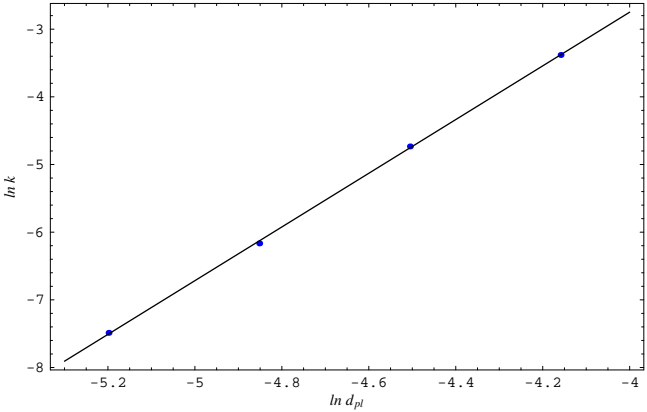}
\label{f:logkvslogdpl}}
\subfigure[$k$ vs.~$m^2/\tau_f^5$ and best fit line $1.9\,m^2/\tau_f^5-2.5\times 10^{-11}$]{
\hspace{-0.1\textwidth}
\includegraphics[width = 0.597\textwidth]{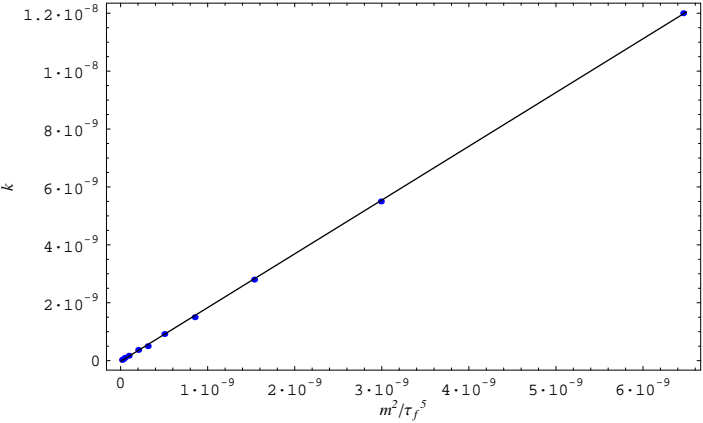}
\label{f:kvsm2tauf5}}
\caption[]{The relationship between the diffusion parameter and underlying model parameters.}
\end{center}
\end{figure}

Plotting $\ln k$ vs.~$\ln \tau_f$, Figure~\ref{f:logkvslogtauf}, for all 11 values of $\tau_f$ reveals the dependence $k\sim\tau_f^{-5}$. 

Dimensional analysis now allows the dependence on the final parameter, $d_{pl}$, to be determined, but for completeness it was checked through simulations. For four different causal set sizes $N=\{4096,\,8192,\,16384,\,32768\}$, 500 trajectories were evolved with fixed $\tau_f$ and $m$. Again, best fit values of $k$ were determined for each value of $N$. Figure~\ref{f:logkvslogdpl} clearly shows, despite only four data points, the dependence $k\sim d_{pl}^4$. Thus, $k\sim m^2 d_{pl}^4/\tau_f^5$ for the swerves model. The constant of proportionality can of course be determined: returning to the varying $\tau_f$ data and working in discreteness units where $d_{pl}=1$ it is found that $k\approx2m^2/\tau_f^5$, as shown in Figure~\ref{f:kvsm2tauf5}.

This relationship will be crucial if and when there is an independent reason for a particular value of 
$\tau_f$. For example, it has been hypothesized that causal set theory will contain a `nonlocality' scale~\cite{Sorkin:2007qi, Sorkin:2009bp}. The forgetting parameter that appears in the above particle models may be a measure of this nonlocality scale. Sorkin~\cite{Sorkin:2007qi} estimates the nonlocality scale to be of the order $10^{-12}cm$, or $10^{20}$ in Planck units. If the results above are extrapolated to such large values of $\tau_f$, a proton with $m=10^{-20}$ would have a diffusion parameter of order $k\sim 10^{-140}$. The diffusion parameter for hydrogen molecules (the factor of two mass difference is inconsequential here) has been constrained to be $k<10^{-102}$~\cite{Dowker:2003hb}. The above estimate is thus not ruled out by existing constraints, but is also, unfortunately, too small to be currently tested.

\section{Conclusions}
There is a considerable focus in quantum gravity phenomenology on violations of Lorentz invariance despite there being no evidence to date that Lorentz invariance is violated. Causal set theory offers a way to investigate Lorentz invariant quantum gravity phenomenology. Earlier work led to a diffusion equation describing the behaviour of massive particles in a discrete spacetime in the continuum limit. Although some underlying models were proposed, no formal connection between the models and the continuum behaviour was made and it was expected that the limitations on the size of causal sets that can be simulated would prohibit any direct demonstration of the diffusion behaviour. The results given here show that microscopic models of particle motion do indeed give rise to diffusion and, moreover, it is not necessary to take the discreteness scale, $d_{pl}$, to zero for this behaviour to occur. The models discussed here are classical point particle models, and there is no claim that they are the true description of particles in causal set theory. This work demonstrates, however, that we can gain much useful information about the observable consequences of discreteness through simulations, despite computational limitations on causal set size.

\ack
The author would like to thank Fay Dowker for ongoing support, Steven Johnston for helpful discussions, and David Rideout for supplying and providing support for the CausalSets toolkit. This work was supported by Tertiary Education Commission of New Zealand (TAD1939) and the Fund for Women Graduates.
\bibliographystyle{unsrt}
\bibliography{../BibliographyMasterCopy/refs} 
\end{document}